\documentclass[reprint, superscriptaddress,
amsmath,amssymb,
aps,
prl,
 floatfix,
]{revtex4-1}

\usepackage[normalem]{ulem} % Must be deleted before submission

\usepackage[colorlinks=true,citecolor=blue,linkcolor=blue,urlcolor=blue]{hyperref}
\usepackage{prlfonts}
\usepackage{graphicx}% Include figure files
\usepackage{dcolumn}% Align table columns on decimal point
\usepackage{bm}% bold math
\usepackage[dvipsnames]{xcolor}
\definecolor{red}{RGB}{250, 0, 0}
\usepackage{amsmath}
\usepackage{mathtools}
\usepackage{color,soul}

\newcommand{\apjl}{Astrophys.\ J. Lett.}
\newcommand{\aap}{Astron.\ Astrophys.}

\newcommand{\ssr}{Space Sci. Rev.}
\begin{document}

%\preprint{APS/123-QED}

\title{High-accuracy Measurements of Core-excited Transitions in Light Li-like Ions}

\author{Moto Togawa}
\affiliation{Max-Planck-Institut f\"ur Kernphysik, Saupfercheckweg 1, 69117 Heidelberg, Germany}%
\affiliation{European XFEL, Holzkoppel 4, 22869 Schenefeld, Germany}%
\affiliation{Heidelberg Graduate School for Physics, Ruprecht-Karls-Universität Heidelberg,
Im Neuenheimer Feld 226, 69120 Heidelberg, Germany}%
\email{togawa@mpi-hd.mpg.de}

\author{Steffen K\"uhn}%
\affiliation{Max-Planck-Institut f\"ur Kernphysik, Saupfercheckweg 1, 69117 Heidelberg, Germany}%
%\email{steffen.kuehn@mpi-hd.mpg.de}

\author{Chintan Shah}%
\affiliation{NASA/Goddard Space Flight Center, 8800 Greenbelt Road, Greenbelt, MD 20771, USA}%
\affiliation{Center for Space Sciences and Technology, University of Maryland, Baltimore County, 1000 Hilltop Circle, Baltimore, MD 21250, USA }
\affiliation{Max-Planck-Institut f\"ur Kernphysik, Saupfercheckweg 1, 69117 Heidelberg, Germany}%
%\email{chintan.shah@mpi-hd.mpg.de}

\author{Vladimir A.~Zaytsev}
\affiliation{Max-Planck-Institut f\"ur Kernphysik, Saupfercheckweg 1, 69117 Heidelberg, Germany}%
%\email{v.a.zaytsev@spbu.ru}

\author{Natalia S.~Oreshkina}
\affiliation{Max-Planck-Institut f\"ur Kernphysik, Saupfercheckweg 1, 69117 Heidelberg, Germany}%
%\email{natalia.oreshkina@mpi-hd.mpg.de}

\author{Jens Buck}%
\affiliation{Institut für Experimentelle und Angewandte Physik, Christian-Albrechts-Universität zu Kiel, Kiel, Germany}%
%\email{jens.buck@desy.de}

\author{Sonja Bernitt}%
\affiliation{GSI Helmholtzzentrum für Schwerionenforschung, Planckstraße 1, 64291 Darmstadt, Germany}%
\affiliation{Helmholtz Institute Jena, Fröbelstieg 3, 07743 Jena, Germany}
\affiliation{Institute for Optics and Quantum Electronics, Friedrich Schiller University Max-Wien-Platz 1, 07743 Jena, Germany}
\affiliation{Max-Planck-Institut f\"ur Kernphysik, Saupfercheckweg 1, 69117 Heidelberg, Germany}%
%\email{s.bernitt@hi-jena.gsi.de}

\author{René Steinbrügge}%
\affiliation{Deutsches Elektronen-Synchrotron (DESY), Notkestrasse 85, 22607 Hamburg, Germany}
%\email{rene.steinbruegge@desy.de}

\author{J\"orn Seltmann}%
\affiliation{Deutsches Elektronen-Synchrotron (DESY), Notkestrasse 85, 22607 Hamburg, Germany}
%\email{joern.seltmann@desy.de}

\author{Moritz Hoesch}
\affiliation{Deutsches Elektronen-Synchrotron (DESY), Notkestrasse 85, 22607 Hamburg, Germany}
%\email{moritz.hoesch@desy.de}

\author{Christoph H. Keitel}
\affiliation{Max-Planck-Institut f\"ur Kernphysik, Saupfercheckweg 1, 69117 Heidelberg, Germany}%
%\email{christoph.keitel@mpi-hd.mpg.de}

% \author{Michael Meyer}
% \affiliation{European XFEL, Holzkoppel 4, 22869 Schenefeld, Germany}%

\author{Thomas Pfeifer}
\affiliation{Max-Planck-Institut f\"ur Kernphysik, Saupfercheckweg 1, 69117 Heidelberg, Germany}%
%\email{thomas.pfeifer@mpi-hd.mpg.de}

\author{Maurice A.~Leutenegger}
\affiliation{NASA/Goddard Space Flight Center, 8800 Greenbelt Road, Greenbelt, MD 20771, USA}%
%\email{maurice.a.leutenegger@nasa.gov}

\author{Jos\'e R.~{Crespo L\'opez-Urrutia}}
\affiliation{Max-Planck-Institut f\"ur Kernphysik, Saupfercheckweg 1, 69117 Heidelberg, Germany}%
\email{crespojr@mpi-hd.mpg.de}

\date{\today}% It is always \today, today,

\begin{abstract}
% Two $1s$-core-excited soft x-ray transitions (dubbed \textit{q} and \textit{r}) from $1s^2 2s ^1P_{1/2}$ to the respective upper levels $1s(^{2}S)2s2p(^{3}P) ^{2}P_{3/2}$ and $^{2}P_{1/2}$ of Li-like oxygen, fluorine and neon were for the first time resolved and calibrated using several nearby transitions of He-like ions. The source of a key uncertainty causing systematic fluctuations generally affecting the accuracy of photon energies at soft x-ray monochromators is identified and suppressed by means of photoelectron spectroscopy leading to 10 fold jump in accuracy compared to previous studies.  This allowed us to verify the agreement of previous experiment and theory at a precision of 5ppm, close to that of current Li-like theory.\\
% One of the here studied line, an essential diagnostic in the study of outflow velocities of AGNs

The transition energies of the two $1s$-core-excited soft X-ray lines (dubbed \textit{q} and \textit{r}) from $1s^2 2s ^1S_{1/2}$ to the respective upper levels $1s(^{2}S)2s2p(^{3}P) ^{2}P_{3/2}$ and $^{2}P_{1/2}$ of Li-like oxygen, fluorine and neon were measured and calibrated using several nearby transitions of He-like ions. The major remaining source of energy uncertainties in monochromators, the periodic fluctuations produced by imperfect angular encoder calibration, is addressed by a simultaneously running photoelectron spectroscopy measurement. This leads to an improved energy determination of 5 parts per million, showing fair agreement with previous theories as well as with our own, involving a complete treatment of the autoionizing states studied here. Our experimental results translate to an uncertainty of only 1.6\,km/s for the oxygen line \textit{qr}-blend used to determine the outflow velocities of active galactic nuclei, ten times smaller than previously possible.

% %including our own that account for the dominant electronic correlation effects as well as contributions from quantum electrodynamics, nuclear size, and normal as well as specific relativistic mass shifts. 
% {\red , which applies a more rigorous, complete treatment of the here studied autoionizing states.}

% %This leads to improved energy determination at 5 parts per million, which agree with available theory as well as with our own new calculations, which account for the dominant electronic correlation effects as well as contributions from quantum electrodynamics, nuclear size, and normal as well as specific relativistic mass shifts. 
% {\red , which applies a more rigorous, complete treatment of the here studied autoionizing states.}
% This translates for the oxygen line \textit{qr}-blend used for determination of outflow velocities of active galactic nuclei into an uncertainty of only 1.6\,km/s, ten times lower than currently possible.%The present benchmark of correlations in light three-electron ions is helpful for the further development of theory of few-electron systems.
\end{abstract}

\maketitle

%%---CS----
%%---astro/plasma/motovation---
High-resolution grating spectrometers onboard the Earth-orbiting X-ray telescopes Chandra and XMM-Newton have enabled the detection of inner-shell X-ray absorption lines in ionized outflows of active galactic nuclei (AGN) and the neutral interstellar medium (ISM) and warm-hot intergalactic medium (WHIM)~\cite{Kaspi_2001, Behar_2002_ApJ, Lee_2001,  Richter2008}. They reveal the physical conditions of the ionized absorbing medium, among others the velocities of outflows, plasma densities and temperatures~\cite{Behar_2003,Steenbrugge_2003,Blustin_2002}. 
The strongest $1s-2p$ inner-shell absorption lines of light Li-like ions (referred to as \textit{q}, \textit{r}, following notation of Gabriel) are among the most important lines observed in such environments.
%Therefore, the lines used for such investigation should be prominently present in astrophysical plasma and their underlying electronic structure simple enough to allow precise predictions. Ideally, these should be accurately benchmarked by ground based experiments. Good candidates are the strongest $1s-2p$ inner-shell absorption lines of light Li-like ions (now on referred as \text{q}, \text{r}, following notation of Gabriel). Both \text{q} and \text{r} offer radiative decay rates of $\sim 10^{12}$ s$^{-1}$ with an autoionization channel approximately equal in strength.  
However, inaccurate transition energies introduced systematic uncertainties, e.~g., discrepancies of up to 1.3\,eV  \,%1.28\,eV %50\,m\AA \, 
have been seen in predictions of \textit{q} and \textit{r} of oxygen at $\sim$ 560 \, eV. 
This caused a velocity uncertainty of 700 km/s, as large as the outflow velocities in nearby active galaxies, which hampered the understanding of multi-component outflows~\cite{Behar_2002, Holczer2010}. Uncertain transition energies also hindered disentanglement of absorption from different charge states of oxygen in the ISM \cite{Mathur2017}. Experiments at the LLNL EBIT \cite{Schmidt_2004} reduced the uncertainty to 20--40 km/s, but individual core-excited fine-structure levels were not resolved. The \textit{q} and \textit{r} lines of Ne$^{7+}$ have also been proposed as electron-density diagnostics in flares of stellar coronae, but suffer both from their overlap with L-shell lines from iron and theoretical uncertainties \cite{Wargelin_2001}.
% \sout{Nonetheless, their measured values revealed deviations as substantial as 1100 km/s compared to different theoretical predictions~\cite{Schmidt_2004,Gu_2005}.}
%\textbf{({\red{MT: Cannot find Wargelin1995 @CS Do you have a doi? https://journals.aps.org/pra/pdf/10.1103/PhysRevA.63.022710}}Add wargelin et al 1995, Ne qr at 43 km/s; )}
%\\
Similar problems occur with the X-ray lines \textit{q} and  \textit{r} emitted following dielectronic recombination of helium-like ions needed for determining electron densities and temperatures in magnetically confined fusion plasmas \cite{Marchuk_2004,Beiersdorfer_2015,Wargelin_2001}.% \textbf{(add wargelin1995 here too)}
The utility the aforementioned cases depend on the quality of the utilized atomic data.

%---theory/QED/improvement motivation---
Unfortunately, Li-like ions still challenge contemporary calculations, which have uncertainties at the few-meV level for core-excited levels \cite{yerokhin_2017,yerokhin_2017_erratum,yerokhin_2012,zaytsev_2020}, far greater than in H-like and He-like systems \cite{yerokhin_2019,vainshtein_1985}. 
With few exceptions \cite{Machado_2020,Schlesser_2013}, earlier X-ray measurements could not benchmark predictions~\cite{Azarov_2022} of the core-excited states ($1s(^{2}S)2s2p(^{3}P) ^{2}P_{3/2}$ and $^{2}P_{1/2}$) due to limited resolution \cite{mannervik_1997,kramida_2006,Schmidt_2004,Gu_2005}. %
Machado \textit{et al.}\cite{Machado_2020} and Schlesser \textit{et al.} \cite{Schlesser_2013} used for \textit{q} and \textit{r} in Ar and S a crystal spectrometer with a resolving power of $\sim$12000, achieving a wavelength accuracy better than 2-5 ppm, but at energies below 1\,keV available crystals and gratings offer neither enough reflectivity nor resolving power. %Examples. Peter's KAP and RAP crystal spectrometers. (The resolving power of these crystal spectrometers is 700 in first order and 1300-1500 in second order.)}
%
% !Intrashell transitions are not our main focus!: Bound-state QED contributions in intra-L-shell $2s_{1/2} - 2p_{3/2,1/2}$ transitions of Li-like ions were studied in traps~\cite{beiersdorfer_1993} and accelerator facilities \cite{brandau_2008,brandau_2003,feili_2000,bosselmann_1999}.  
%
Other experiments using merged beams of ions and photons \cite{mclaughlin_2017} were limited by insufficiently accurate calibrations with molecular X-ray absorption spectra \cite{Bizau_2015,coreno_1999}. %These references usually associated with large uncertainties, which are among other things caused by their broad structure due to many autoionization channels, as well as rovibrational features that complicated the energy calibration.
Systematic uncertainties were recently substantially reduced down to 15\,ppm by using $1s-np$ transitions in He-like ions to independently re-calibrate molecular transitions \cite{leutenegger_2020, stierhof_2022}.
%
%Recently, it was demonstrated the feasibility of achieving a notably high spectral resolution, exceeding 20\,000, in the soft X-ray region by using high-resolution monochromators~\cite{Hoesch_2022}. 
State-of-the-art soft X-ray monochromators with resolution exceeding 20\,000  \cite{Hoesch_2022} still suffer from calibration problems  resulting from periodic errors of angular encoders in use ~\cite{kuehn_2022}.

In this Letter, we present measurements at the P04 beamline of the PETRA-III storage ring in Hamburg, Germany \cite{viefhaus_2013} resolving the $q$ and $r$ lines in Li-like oxygen, fluorine, and neon, and yielding transition energies with an accuracy of approximately 5\,ppm. 
A more than 10-fold improvement made possible by mitigating
systematic fluctuations of angular encoders by simultaneous high-resolution photoelectron spectroscopy (XPS). 
%{\red{(unc. 3.4 meV at 563 eV (quad); unc. 4.8 meV at 563 eV (lin)}}.
%resolving power of 20000 at Ne $q$ and 30000 at O $q$. 
%
% This improves fourteen-fold earlier work \cite{Schmidt_2004}. The measured rest wavelengths have an uncertainty of $\sim$1.6\,km/s, much lower than the 50--150\,km/s values \cite{Behar_2002,Schmidt_2004,Bozzo_2023} of Chandra \cite{Weisskopf_2000} and XMM-Newton \cite{jansen_2001,denHerder_2001} observations, and the projected statistical uncertainties of $\sim$33\,km/s of recently launched XRISM observatory~\cite{xrism}. 
% %
Our accurate measurements benchmark predictions accounting for electron-electron correlation, QED and nuclear size effects, as well as general and specific relativistic mass shifts. Moreover, we provide accompanying large-scale calculations, which exhibit theoretical uncertainties on par with state-of-the-art predictions while taking the effect of autoionization shifts into account ~\cite{zaytsev_2020}.
We thereby provide, both experiment and theory to test and benchmark the state-of-the-art calculations of Li-like theory for few selected elements.

\begin{figure}
    \centering
    \includegraphics[width=\columnwidth]{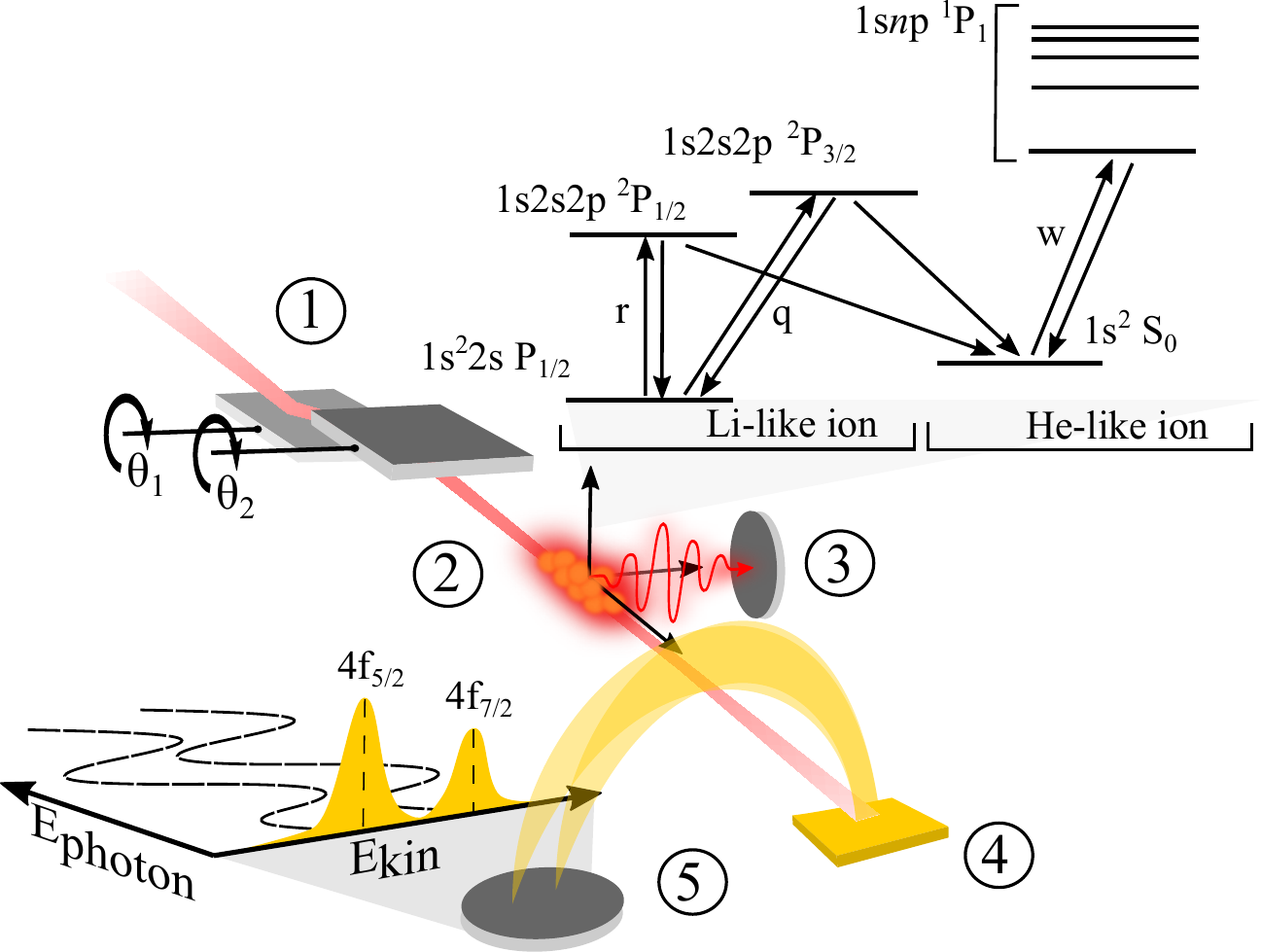}
    \caption{Scheme of our experiment. Passing a monochromator (1) photons are focused onto the trapped HCI (2), with the relevant transitions and energy levels indicated. Fluorescence photons following resonant excitation due to the incident photon beam are recorded by a SDD (3). The outgoing beam continuously generates photoelectrons from a gold target (4). A PES resolves the two prominent $4f$ lines on a detector (5) to monitor energy fluctuations. }

    \label{fig:Grotrian}
\end{figure}
% At P04 \cite{viefhaus_2013}, an undulator generates a beam of up to $10^{12}$ photons per second {\red{Jörn:  Undulator emits $1e15 ph/s/0.1 B.W.$ We get $10^{12}$ at the fokus}}. 
We acquire absorption spectra by scanning the incident photonenergy while counting the number of fluorescence photons after resonant excitation, which are detected by silicon-drift detectors (SDD).
At P04 \cite{viefhaus_2013}, a photon beam of $10^{14} \gamma/s$ at $0.1\%$ bandwidth is generated by an undulator, which after monochromatization and transport losses results in an approximately $10^{11} \gamma/s$ flux at the focus, which is placed at center of our compact electron beam ion trap, PolarX-EBIT \cite{micke_2018}.
The ions are produced by PolarX-EBIT, by injecting a tenuous atomic or molecular beam containing the element of interest. It crosses the electron beam, which is focused by a magnetic field and set to an energy sufficient to generate and trap the respective He-like and Li-like ions, but below their K-shell excitation threshold. This ensures a good signal-to-noise in the silicon-drift detectors, which are equipped with 500 \, nm aluminium filters for blocking most of the low energy stray light. Both detectors are mounted side-on for registering soft fluorescence X-rays produced by electron impact and, crucially, upon resonant photoexcitation. This method was demonstrated at the free-electron laser FLASH as soft X-ray laser spectroscopy \cite{Epp_2007} and later also applied at synchrotron-radiation facilities \cite{Simon_2010,Rudolph_2013,Steinbruegge_2015,leutenegger_2020,kuehn_2020,kuehn_2022}.

After optimizing the monochromator using the strong, narrow \textit{w} ($1s2p \, ^{1}P_{1}$ to $1s^{2} \, ^{1}S_{0}$) line of oxygen \cite{Hoesch_2022} to achieve a resolving power of more than 30\,000, we can resolve \textit{q} and \textit{r}. We then scan several times in discrete monochromator steps of nominal energies a range containing the \textit{q} and \textit{r} transitions of the Li-like ion, and calibration lines including the \textit{w} line of the He-like ion of the same element, as well as a short series of $1s^{2}$ to  $1snp$ transitions of the next lower element in atomic number.
\begin{figure*}[ht]
    \centering
    \includegraphics[width=0.95\textwidth]{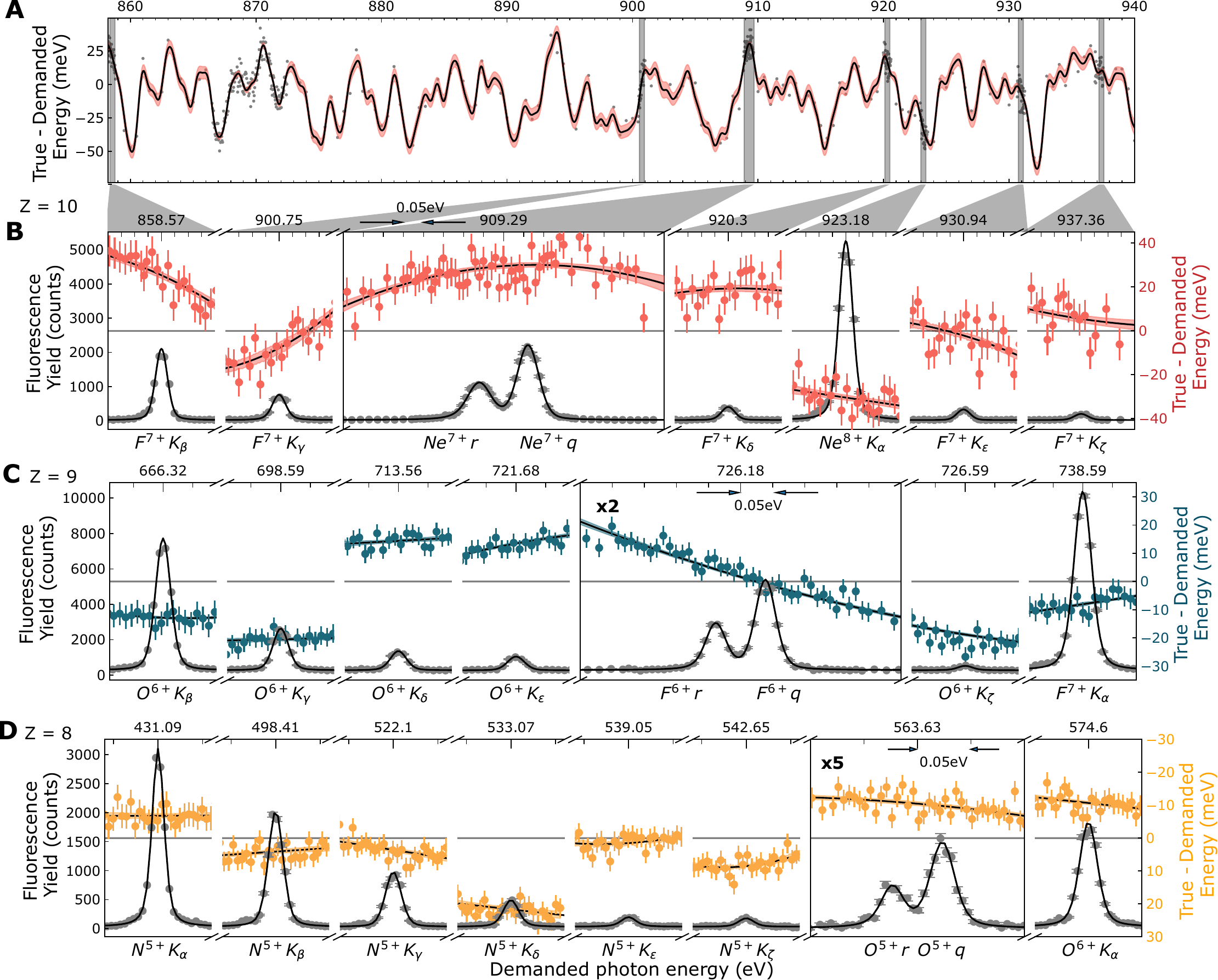}
    \caption{
    (A) Trace of energy deviations recorded by means of XPS of Au 4f$_{5/2}$ photoelectrons. The trace covers the energy range of the Li-like neon \textit{q} and \textit{r} measurement. (B,C,D) Example scans of \textit{q} and \textit{r} lines for neon, fluorine and oxygen and respective calibration lines are superimposed with their, simultaneously acquired, photoelectron trace, which monitors the deviations of the demanded photon energy from the actual energy.
    Each scan contains the resolved \textit{q} and \textit{r} transitions of the Li-like ion, as well as various He-like calibration lines.}
    \label{fig:overview}
\end{figure*}

Knowing the actual photon energy depends on accurate readings of grating and mirror angles ($\theta_1, \theta_2$ in Fig. \ref{fig:Grotrian}) in the monochromator. Angular encoders measure them by the transmission of light between several glass disks patterned with opaque marks that overlap only at certain rotation-angle steps. While the narrow line width of the recorded transitions calls for steps of roughly $4 \times 10^{-5}$ degree, the spacing between the 36\,000 reference marks of the Heidenhain RON 905 encoders used in the P04 monochromator delivers only $10^{-2}$-degree dark-bright cycles. Between those marks, encoders interpolate the angle changes thousand-fold using the quadrature signals of several diodes measuring the light modulated by minor disk rotations. This procedure is very sensitive to imperfections of those analog signals, and thus empirical look-up tables have to be regularly generated and stored in the hardware. However, residual errors remain. Previous observations at P04 and beamlines elsewhere showed that nominal photon energies calculated from such interpolated readouts had periodic sub-divisional errors \cite{Krepansky_2011,follath_2010}, leading to fluctuations in the photon-energy readout. With two encoders needed to measure the diffraction angle, a double interpolation uncertainty should affect most X-ray monochromators worldwide \cite{Krepansky_2011,follath_2010}. As we will discuss in the following, periodic changes with peak-to-peak amplitudes of up to 70\,meV have been found in our case.

Since the off-axis electron gun of PolarX-EBIT lets the photon beam exit downstream unimpeded, we perform XPS \cite{Siegbahn_1957} measurements for monitoring fluctuations of the actual photon-beam energy (See Fig. \ref{fig:Grotrian}). The hemispherical photoelectron spectrometer (PES), ASPHERE \cite{asphere_2021} is permanently installed at P04 several meters downstream of the open port where PolarX-EBIT is mounted. After passing through it, the photon beam illuminates a gold target mounted in PES that is electrostatically biased, where photoelectrons are emitted from the Au $4f_{5/2,7/2}$ states known for their large cross sections \cite{Siegbahn_1957,seah_1998,aksela_2012} %(binding energies $E_{4f}\approx$\,80\,eV) 
with a kinetic energy given by $E_{kin} = E_{\gamma} - E_{4f} + V_{bias}$. While $V_{bias}$ is scanned in par with the nominal photon energy in order to keep $E_{kin}$ nominally constant, PES selects $4f_{5/2,7/2}$ photoelectrons within a narrow ($\approx$15\,eV) range encompassing both states, and guides them to a microchannel-plate-amplified phosphor screen imaged on a camera. This high selectivity together with the short term stability of ASPHERE allows us to monitor periodic energy fluctuations from the nominal, linearly growing energy of each scan. 
%Spectra are acquired at approximately 500\,meV FWHM and count rates of $\sim10^4$.
\begin{figure*}[ht]%[htbp]
    \centering
    \includegraphics[width=\textwidth]{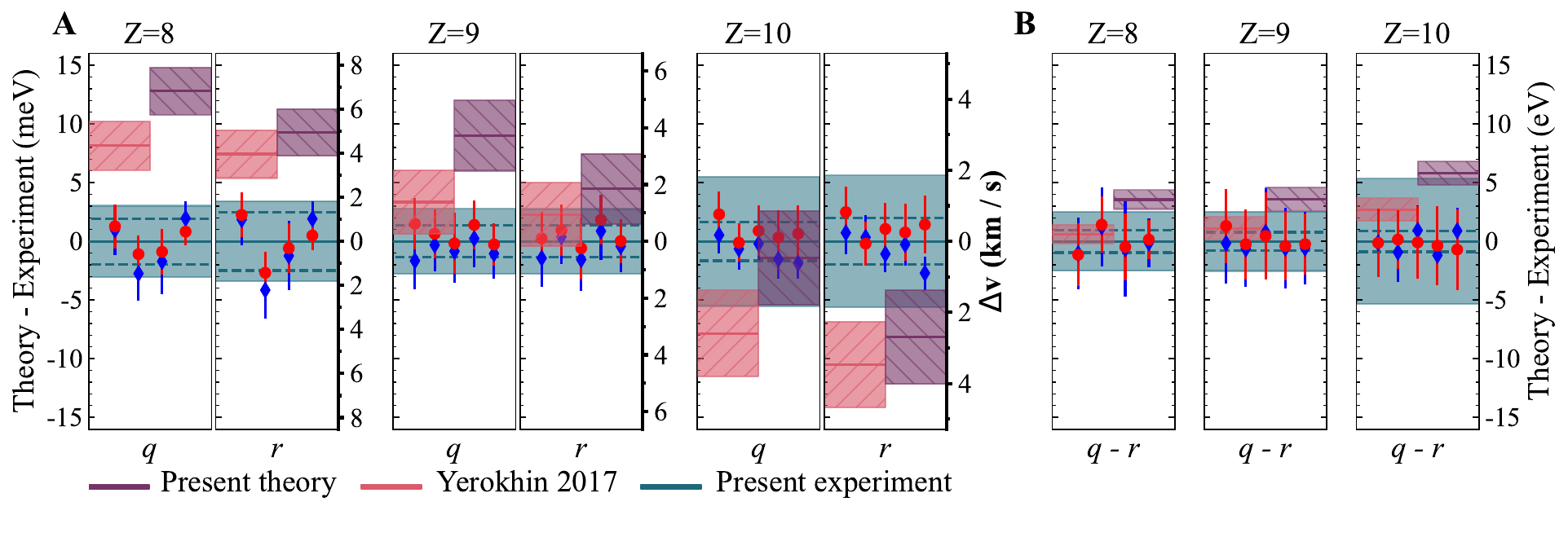}
    \caption{Comparison of our results with theory from Refs.~\cite{yerokhin_2017,yerokhin_2017_erratum}. Theory-experiment energy difference of \textit{q} and \textit{r} (A) and their fine-structure splitting (B). Dashed lines mark experimental 1-sigma uncertainties, excluding those of our XPS data. Area shaded in green includes all uncertainties. Predictions and uncertainties of Refs.~\cite{yerokhin_2017,yerokhin_2017_erratum} as well as of our calculations are shaded in red and purple, respectively. The bold right-side axes in sub-figure (A) show the accuracy in units of km/s, corresponding to the uncertainty of the AGN-outflow velocity.}
    \label{fig:comparison}
\end{figure*}
While the $4f_{5/2,7/2}$ photoelectron peaks should have fixed positions at the detector, actual energy fluctuations induce small centroid shifts of both peaks. After projection of the detector image onto its dispersive axis, we continuously monitor them with a few meV statistical uncertainty by fitting two Voigt peaks and a linear background. Alternative fit models did not significantly improve the fits. During photon energy scans, the $4f_{5/2,7/2}$ peaks' oscillations (see Fig.\ref{fig:overview} A) reflect the interpolation inaccuracies of the two angular encoders. Two distinct oscillation periods arise from the different distances of mirror and grating to undulator X-ray source and exit slit. This recording yields the photoelectron traces used for correction.
To calibrate the kinetic-energy range covered by these traces on the photoelectron detector image, we scan the bias voltage at a constant photon energy, shifting the  $4f_{5/2,7/2}$ peaks across the detector. Subsequently, traces are locally modeled at resonances using low-degree polynomials within narrow energy windows (See Fig. \ref{fig:overview}), and added as corrections to the nominal, yet uncalibrated photon energy scale derived from the monochromator-angle readout. Using this modified scale, the centroids of the HCI fluorescence resonances are determined by fitting Voigt functions. %Their Gaussian component is mainly associated to thedepends on the ion-temperature-dependent Doppler-broadened line profile, plus instrumental effects such as vibrations of the optical components at the $\mu$m level. 
Under present experimental conditions, we see several sources comprising Gaussian and Lorentzian components, respectively. We associate the Gaussian contribution with the inherent limitations in resolution of the monochromator and the thermal motion of the ions. The Lorentzian width, as shown in~\cite{kuehn_2022}, stems from the finite lifetime of the excited levels  and the pseudo-Lorentz instrumental component due to X-ray diffraction at beamline components~\cite{follath_2010}. 
% Under present experimental conditions, we estimate the Doppler Broadening ($\approx$ 10\, meV) to be the leading contribution to the observed Gaussian width. Following closely behind is the instrumental energy profile of the incident beam.
% We found that the Lorentzian component of the Voigt functions is not solely due to the different natural linewidths, but also contains a recognizable contribution from the incident beam. 
For absolute calibration of the photon energy in each scan, we assign to the measured positions of the He-like transitions predicted energy values from Ref.~\cite{yerokhin_2019}, and fit the corresponding dispersion curves using linear functions, except for \textit{q} and \textit{r} of Li-like oxygen, where a second-order polynomial was needed.

We found a systematic shift in the energies of \textit{q} and \textit{r} depending on whether the correction was derived from the Au $4f_{7/2}$ or the $4f_{5/2}$ peak (blue and red data points in Fig.~\ref{fig:comparison}).  By taking a weighted average of these individual results, we find this shift being largest for neon with approximately 1.8\,meV, for fluorine 1\,meV 
and negligible for oxygen measurements, and take it into account with an accordingly enlarged systematic uncertainty. These uncertainty bands are depicted as dashed lines in Fig.~\ref{fig:comparison}. Repeated XPS measurements also revealed a broader distribution of  $4f_{7/2}$, $4f_{5/2}$ centroids than statistically expected, which we attribute to instabilities in the voltage sources of PES. %We estimate this systematic error from the distribution widths found (Ne: 5.3 meV, F: 2.4 meV, O: 2.3 meV), and add it in quadrature to the total (see supplemental material for details).
We estimate this systematic error from the distribution widths found respectively as 5.3\,meV, 2.4\,meV and 2.3\,meV for Ne, F, and O and add it in quadrature to the total. See Supplemental Material \cite{Supplemental} for details on error estimation.

We then compare the measured energies of \textit{q} and \textit{r} with high-precision calculations of both the ground state and the excited $1s2s2p$ states, including contributions from electronic correlations, quantum electrodynamics (QED), and nuclear recoil. Since the excited states can decay via electron emission, a so-called Auger-Meitner channel (see Fig. \ref{fig:Grotrian}), they do not have square integrable wave functions. For this reason, the energies of autoionizing states can exhibit a strong dependence on the basis set parameters, which limits the accuracy of the standard high precision approaches such as CI or Coupled Cluster. To properly account for the energy shift resulting from the Auger-Meitner channel~\cite{zaytsev_2019}, we have used the complex scaling method~\cite{ho_1983, moisev_1998, lindroth_2012,zaytsev_2020} to evaluate the energies of $1s2l2l'$ levels of Li-like oxygen, fluorine, and neon with extended configuration space.

In Table~\ref{table:Results} and Fig.~\ref{fig:comparison}, we compare the experimental data and our new calculations with the existing predictions of~\citet{yerokhin_2017} based on the basis-balancing method for the treatment of the autoionization channel. For all elements, our theoretical values for the \textit{q} and \textit{r} transitions show a shift of $\sim$5\,meV and $\sim$2\,meV, respectively, from those of Ref. \cite{yerokhin_2017,yerokhin_2017_erratum}. Although our complex rotation method is better suited for the complete treatment of states with autoionization channels, our experiment shows overall better agreement with Ref.~\cite{yerokhin_2017}. It is interesting to note that both predictions of fine structure splitting, i.e. the differences between $q$ and $r$, agree well with our experiment. This suggests that the likely cause of the discrepancy observed in the absolute energy comparison may be due to electron correlation effects rather than the QED corrections. 

\begin{figure}
    \centering
    \includegraphics[width=0.5\textwidth]{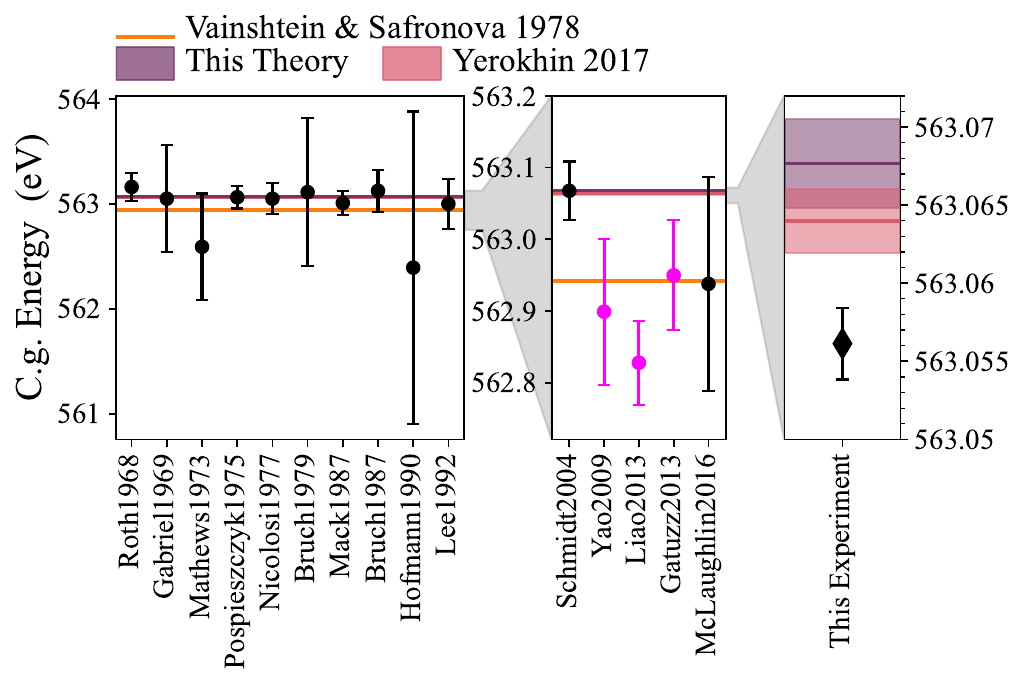}
    \caption{Experimental values for the center-of-gravity energy of the blended \textit{q}\textit{r} line of Li-like oxygen (\cite{Azarov_2022,roth_1968,gabriel_1969,matthews_1973,pospieszczyk_1975,Nicolosi_1977,Bruch_1979,mack_1987,bruch_1987,hofmann_1990,Lee_1992,Schmidt_2004,Yao_2009,Liao_2013,Gatuzz_2013,Mclaughlin_2016}). Astrophysical observations (magenta circles); predictions and their uncertainties: Ref.~\cite{yerokhin_2017,yerokhin_2017_erratum} (red); this work (purple); Ref.~\cite{Vainshtein_1978} (orange).
    }
    \label{fig:r_comparison}
\end{figure}

Fig.~\ref{fig:r_comparison} compares our results with predictions and measurements of the unresolved oxygen \textit{qr}-blend of other works. We also include earlier predictions from~\citet{Vainshtein_1978} showing one significant digit more than other theoretical works \cite{Gabriel_1972,Behar_2002,Pradhan_2003}. Interestingly, the center-of-gravity value of 562.9419\,eV by~\citet{Vainshtein_1978} agrees better with astrophysical observations (\cite{Yao_2009,Liao_2013,Gatuzz_2013}) than other laboratory measurements. However, both our theoretical and experimental results align more favorably with Refs.~\cite{yerokhin_2017,yerokhin_2017_erratum}.

\begin{center}
\begin{table}
\renewcommand{\arraystretch}{1.2}
% Error of CoG values updated on 27.05.24
\caption{Measured energies of the $1s2s2p \,\, {}^2P_{3/2}$ (\textit{q}) and $1s2s2p \,\, {}^2P_{1/2}$ (\textit{r}) transitions, derived center-of-gravity (c.g.) and differences from predictions. All values are given in eV} %\vspace{0.2cm}
\begin{tabular}{ l c  r r r r} 
\hline \hline
Element &  & \multicolumn{2}{c}{This work} & Refs.~\cite{yerokhin_2017,yerokhin_2017_erratum} & \; \\ %\cite{zaytsev_2020}
        &  &        Exp.       &    Theory & Theory &\\
\hline
 {Z = 8}  & \textit{q}     & 563.0712(30)   & 563.084(2) & 563.079(2)   \\ 
          & \textit{r}     & 563.0257(34)   & 563.035(2) & 563.033(2)   \\
          & c.g. & 563.0560(23) & & 563.064(2) \\
          & \textit{q - r} &  \phantom{00} 0.0456(25) & \phantom{00}0.0492(8)  &   \phantom{00}0.0463(8) \\ 
 {Z = 9}  & \textit{q}     & 725.3720(28)   & 725.381(3) & 725.375(3)  \\ 
          & \textit{r}     & 725.2945(28)  & 725.299(3) & 725.297(3)  \\ 
           & c.g. & 725.3462(21) & & 725.349(3)\\
          & \textit{q - r} &\phantom{00} 0.0774(25)  &   \phantom{00}0.081(1) &   \phantom{00}0.079(1)   \\ 
 {Z = 10} & \textit{q}     & 908.2019(55)   & 908.200(4) & 908.194(4)  \\ 
          & \textit{r}     & 908.0796(57)   & 908.071(4) & 908.069(4)  \\ 
           & c.g. & 908.1607(41) & & 908.151(4) \\
          & \textit{q - r} &\phantom{00} 0.1222(53) &   \phantom{00}0.128(1) &   \phantom{00}0.125(1)   \\ 
\hline
\end{tabular}
\label{table:Results}
\end{table}
\end{center}
\indent

By combining soft
X-ray laser spectroscopy of accurately \textit{ab-initio}-predicted narrow He-like transitions with synchronous XPS measurements, we eliminate encoder interpolation errors generally affecting energy determinations with monochromators, and solve a long-standing problem of such devices.
Thus, our soft X-ray energy measurements below 1 keV are the most accurate to date. For the studied low-$Z$ elements, electronic correlations are dominant.
Nonetheless, QED effects in these autoionizing systems cause shifts  carrying theoretical uncertainties as large as those of Dirac-Coulomb-Breit terms. 
%Our benchmarking experiment establishes a bound for the theoretical uncertainty in correlation-dominated low-$Z$ systems.
Understanding these enables more robust tests of QED theory and mass-shift contributions in strong fields using heavier ions, where correlation effects become smaller. 
Furthermore, our results re-calibrate earlier works, and immediately benefit \textit{XRISM} \cite{xrism}, a recently launched X-ray observatory furnished with a high-resolution X-ray microcalorimeter. Additionally, our data provides accurate reference lines, which allow full utilization of upcoming X-ray observatories such as \textit{Athena}~\cite{pajot2018athena} and \textit{Arcus}~\cite{arcus_instrum}, which have targeted resolving power of 1000-3500 and uncertainty of below 10\,km/s, call for high accuracy rest wavelength standards of essential soft X-ray transitions.
%-resolution X-ray microcalorimeter and even more upcoming X-ray observatories as \textit{Arcus}, as its targeted resolving power of 3800 and calibration uncertainty of below 105\,km/s urgently calls for high accuracy rest energies for essential soft X-ray transitions. 

\begin{acknowledgments}
Financial support was provided by the Max-Planck-Gesellschaft (MPG) and Bun\-des\-mi\-ni\-ste\-ri\-um f{\"u}r Bildung und Forschung (BMBF) through project 05K13SJ2. C.S. acknowledges support from NASA under award number 80GSFC21M0002 and Max-Planck-Gesellschaft (MPG). M.A.L. acknowledges support from NASA's Astrophysics Program. We acknowledge DESY (Hamburg, Germany), a member of the Helmholtz Association HGF, for the provision of experimental facilities. Parts of this research were carried out at PETRA~III. We thank Jens Viefhaus and Rolf Follath for valuable discussions on X-ray monochromator resolution and performance, and the synchrotron-operation team at PETRA~III for their skillful and reliable work.
\end{acknowledgments}

% \printbibliography
% \bibliographystyle{unsrt}

% \bibliographystyle{ieeetr}
% \bibliography{bibliography}
%\bibliography{bibliography}
%%%%

% apsrev4-2.bst 2019-01-14 (MD) hand-edited version of apsrev4-1.bst
% Control: key (0)
% Control: author (72) initials jnrlst
% Control: editor formatted (1) identically to author
% Control: production of article title (-1) disabled
% Control: page (0) single
% Control: year (1) truncated
% Control: production of eprint (0) enabled

%%%%Below Bibliography 28.05.24
%

\end{document}